\documentclass[aps,prd,twocolumn,
preprintnumbers,
showpacs,floatfix]{revtex4}

\usepackage{bm}
\usepackage{epsfig}
\usepackage{graphics}
\usepackage{psfrag}
\usepackage{amsmath}
\usepackage{textcomp}

\newcommand{\eeq}{\end{equation}}
\newcommand{\beq}{\begin{equation}}
\newcommand{\ba}{\begin{array}}
\newcommand{\ea}{\end{array}}
\newcommand{\bea}{\begin{eqnarray}}
\newcommand{\eea}{\end{eqnarray}}
\newcommand{\vev}[1]{\langle #1\rangle}

\newcommand{\veps}{\varepsilon}

\newcommand{\si}{\sigma}

\newcommand{\gsim}{\mbox{\raisebox{-.9ex}
{~$\stackrel{\mbox{$>$}}{\sim}$~}}}

\newcommand{\pb}{\bar{\phi}}
\newcommand{\Hb}{\bar{H}}
\newcommand{\ka}{\kappa}
\newcommand{\la}{\lambda}

\newcommand{\ten}[1]{\,\cdot 10^{#1}}
\newcommand{\units}[1]{\,\text{#1}}
\newcommand{\mP}{m_{\rm P}}

\DeclareMathOperator{\Tr}{Tr}

\begin{document}

\preprint{UT--STPD--1/08}

\title{Semi-shifted hybrid inflation with
$\rm B-L$ cosmic strings}

\author{George Lazarides}
\email{lazaride@eng.auth.gr}
\author{Iain N.R. Peddie}
\email{inr314159@hotmail.com}
\author{Achilleas Vamvasakis}
\email{avamvasa@gen.auth.gr}
\affiliation{Physics Division, School of
Technology, Aristotle University of
Thessaloniki, Thessaloniki 54124, Greece}

\date{\today}

\begin{abstract}
\par
We discuss a new inflationary scenario which is
realized within the extended supersymmetric
Pati-Salam model which yields an acceptable
$b$-quark mass for universal boundary conditions
and $\mu>0$ by modestly violating Yukawa
unification and leads to new shifted, new smooth,
or standard-smooth hybrid inflation. Inflation
takes place along a ``semi-shifted'' classically
flat direction on which the ${\rm U(1)_{B-L}}$
gauge group remains unbroken. After the end of
inflation, ${\rm U(1)_{B-L}}$ breaks
spontaneously and a network of local cosmic
strings, which contribute a small amount to the
curvature perturbation, is produced. We show
that, in minimal supergravity,
this ``semi-shifted'' inflationary scenario is
compatible with a recent fit to data which uses
field-theory simulations of a local string
network. Taking into account the requirement of
gauge unification, we find that, for spectral
index $n_{\rm s}=1$, the predicted fractional
contribution of strings to the temperature power
spectrum at multipole $\ell=10$ is
$f_{10}\simeq 0.039$. Also, for
$f_{10}=0.10$, which is the best-fit value,
we obtain $n_{\rm s}\simeq 1.0254$. Spectral
indices lower than about $0.98$ are excluded and
blue spectra are slightly favored. Magnetic
monopoles are not formed at the end of
semi-shifted hybrid inflation.
\end{abstract}

\maketitle

\section{Introduction}
\label{sec:intro}

One of the most promising models for inflation
\cite{inflation} (for a review, see e.g.
Ref.~\cite{lectures}) is, undoubtedly, hybrid
inflation \cite{linde}, which is \cite{cop,dss}
naturally realized within supersymmetric (SUSY)
grand unified theory (GUT) models. In the
standard realization of SUSY hybrid inflation,
the spontaneous breaking of the GUT gauge
symmetry takes place at the end of inflation and,
thus, superheavy magnetic monopoles
\cite{monopole} are copiously produced
\cite{smooth1} if they are predicted by this
symmetry breaking. In this case, a cosmological
catastrophe is encountered.

\par
In order to avoid this disaster, one can employ
the smooth \cite{smooth1,smooth2} or shifted
\cite{shift} variants of SUSY hybrid inflation
(for a review, see Ref.~\cite{talks}). In these
inflationary scenarios, which, in their original
realization, are based on non-renormalizable
superpotential terms, the GUT gauge symmetry is
broken to the standard model (SM) gauge group
$G_{\rm SM}$ already during inflation and, thus,
no magnetic monopoles are produced at the
termination of inflation. New versions of these
inflationary schemes can be implemented
\cite{nshift,nsmooth} with only renormalizable
superpotential terms within an extended SUSY
GUT model based on the Pati-Salam (PS) gauge
group $G_{\rm PS}={\rm SU}(4)_c\times
{\rm SU}(2)_{\rm L}\times{\rm SU}(2)_{\rm R}$
\cite{pati}, whose spontaneous breaking to
$G_{\rm SM}$ predicts the existence of doubly
charged \cite{magg} magnetic monopoles. Actually,
this extended SUSY PS model was initially
constructed \cite{quasi}
(see also Ref.~\cite{quasitalks}) for solving a
very different problem. In SUSY models with exact
Yukawa unification \cite{als}, such as the
simplest SUSY PS model (see Ref.~\cite{hw}), and
universal boundary conditions, the predicted
$b$-quark mass is \cite{hall} unacceptably large
for $\mu>0$. However, it can be adequately
reduced if Yukawa unification is moderately
violated. This is achieved by extending the
superfield content of the SUSY PS model so as to
include, among other superfields, an extra pair
of ${\rm SU}(4)_c$ non-singlet
${\rm SU}(2)_{\rm L}$ doublets, which naturally
mix \cite{wetterich} with the main
electroweak doublets.

\par
Fitting the recent data of the Wilkinson
microwave anisotropy probe (WMAP) satellite with
the standard power-law cosmological model with
cold dark matter and a cosmological constant
($\Lambda$CDM), one obtains \cite{wmap} values of
the spectral index $n_{\rm s}$ which are clearly
lower than unity. (Note, though, that some recent
analyses, e.g. Ref.~\cite{newns}, reduce somewhat
the evidence for $n_{\rm s}<1$.) However, in
supergravity (SUGRA) with canonical K\"{a}hler
potential, the above hybrid inflation models
yield \cite{nsmooth,nshift,senoguz} $n_{\rm s}$'s
which are very close to unity or even larger than
it although their running is negligible. This
discrepancy may be resolved
\cite{lofti,king,rehman} by including non-minimal
terms in the K\"{a}hler potential. Alternatively,
if we wish to stick to minimal SUGRA, we can
reduce \cite{mhin} the spectral index predicted
by the hybrid inflationary models by restricting
the number of e-foldings suffered by our present
horizon scale during the hybrid inflation which
generates the observed curvature perturbations.
The additional number of e-foldings required for
solving the horizon and flatness problems of
standard hot big bang cosmology can be provided
by a subsequent second stage of inflation. In
Ref.~\cite{stsm}, we showed that the same
extended SUSY PS model can lead to a two-stage
inflationary scenario yielding acceptable
$n_{\rm s}$'s in minimal SUGRA. The first stage
of inflation, during which the cosmological
scales exit the horizon, is of the standard
hybrid type, while the second stage, which
provides the additional e-foldings, is of the new
smooth hybrid type.

\par
In this paper, we consider an alternative
inflationary scenario which incorporates cosmic
strings \cite{string} (for a textbook
presentation or a review, see e.g.
Ref.~\cite{vilenkin}) and can also be naturally
realized within this extended SUSY PS model with
only renormalizable superpotential terms. As
shown in Ref.~\cite{nsmooth}, in a certain range
of parameters, this model possesses a shifted
classically flat direction along which
${\rm U}(1)_{\rm B-L}$ is unbroken. In order to
distinguish it from the new shifted flat
direction on which $G_{\rm PS}$ is broken to
$G_{\rm SM}$, we will call this flat
direction ``semi-shifted''. This direction
acquires \cite{rc}, as usual, a logarithmic slope
from one-loop radiative corrections which are due
to the SUSY breaking caused by the non-zero
potential energy density on it. So,
it can perfectly be used as an inflationary path
along which ``semi-shifted'' hybrid inflation
takes place. When the system crosses the critical
point at which this path is destabilized, a
waterfall regime occurs during which the
${\rm U}(1)_{\rm B-L}$ gauge symmetry breaks
spontaneously and local cosmic strings are
produced. The resulting string network can then
contribute to the primordial curvature
perturbations.

\par
It has been argued \cite{battye}, that, in the
presence of a small contribution to the curvature
perturbation from cosmic strings, the current
cosmic microwave background (CMB) data can allow
values of the spectral index that are larger than
the ones obtained in the absence of strings.
Therefore, we may hope that our semi-shifted
hybrid inflationary scenario, which does involve
cosmic strings, can be made compatible with the
CMB data even without the use of non-minimal
terms in the K\"{a}hler potential or a subsequent
complementary stage of inflation. Recently, a fit
to the CMB data and the luminous red galaxy data
in the Sloan digital sky survey (SDSS)
\cite{sdss} on large length scales outside the
non-linear regime was performed \cite{bevis1} by
using field-theory simulations \cite{bevis2} of
a dynamical network of local cosmic strings. It
demonstrated that the Harrison-Zeldovich (HZ)
model (i.e. $n_{\rm s}=1$) with a fractional
contribution $f_{10}\approx 0.10$ from cosmic
strings to the temperature power spectrum at
multipole $\ell=10$ is even moderately favored
over the standard power-law model without
strings. For the power-law $\Lambda$CDM
cosmological model with cosmic strings, this fit
yields \cite{bevis1} $n_{\rm s}=0.94-1.06$ and
$f_{10}=0.02-0.18$ at $95\%$ confidence level
(c.l.). We show that, under these
circumstances, the semi-shifted hybrid
inflationary model in minimal SUGRA can easily be
compatible with the data. Note that there is
obviously no formation of PS magnetic monopoles
at the end of the semi-shifted hybrid inflation
and, thus, the corresponding cosmological
catastrophe is avoided.

\par
In Sec.~\ref{sec:semishift}, we summarize the
salient features of the extended SUSY PS model
and sketch the semi-shifted hybrid inflationary
scenario with cosmic strings. In
Sec.~\ref{sec:radcorr}, we calculate the one-loop
radiative correction to the inflationary
potential along the semi-shifted path. In
Sec.~\ref{sec:sugra}, we include the minimal
SUGRA correction to this inflationary potential,
while Secs.~\ref{sec:observables} and
\ref{sec:string} refer to the inflation and
string power spectrum respectively.
Sec.~\ref{sec:numeric} is devoted to the
presentation of our numerical results, which show
that, in minimal SUGRA, our semi-shifted hybrid
inflationary scenario with cosmic strings can
yield a spectral index close to unity and be
compatible with the data. In
Sec.~\ref{sec:gauge}, we discuss briefly gauge
unification. Finally, in
Sec.~\ref{sec:conclusions}, we present our
conclusions.

\section{Semi-shifted hybrid inflation}
\label{sec:semishift}

\par
We consider the extended SUSY PS model of
Ref.~\cite{quasi}, which can lead to a moderate
violation of the asymptotic Yukawa unification
\cite{als} so that, for $\mu>0$, an acceptable
$b$-quark mass is obtained even with universal
boundary conditions. The breaking of $G_{\rm PS}$
to $G_{\rm SM}$ is achieved by the superheavy
vacuum expectation values (VEVs) of the right
handed neutrino type components
of a conjugate pair of Higgs superfields $H^c$
and $\bar{H}^c$ belonging to the $(\bar{4},1,2)$
and $(4,1,2)$ representations of $G_{\rm PS}$
respectively. The model also contains a gauge
singlet $S$ and a conjugate pair of superfields
$\phi$, $\bar{\phi}$ belonging to the (15,1,3)
representation of $G_{\rm PS}$. The superfield
$\phi$ acquires a VEV which breaks $G_{\rm PS}$
to $G_{\rm SM}\times{\rm U(1)_{B-L}}$. In
addition to $G_{\rm PS}$, the model possesses a
$Z_2$ matter parity symmetry and two global
${\rm U}(1)$ symmetries, namely a Peccei-Quinn
and a R symmetry. Such continuous global symmetries
can effectively arise \cite{laz1} from the rich
discrete symmetry groups encountered in many
compactified string theories (see e.g.
Ref.~\cite{laz2}). For details on the full field
content and superpotential, the charge
assignments, and the phenomenological and
cosmological properties of this extended SUSY PS
model, the reader is referred to
Refs.~\cite{quasi,shift} (see also
Ref.~\cite{quasitalks}). This model can lead
to new shifted \cite{nshift} and new smooth
\cite{nsmooth} hybrid inflation based solely on
renormalizable interactions. It can also yield
\cite{stsm} a two-stage inflationary scenario
consisting of hybrid inflation of the standard
type followed  by new smooth hybrid inflation.

The superpotential terms which are relevant for
inflation are
\beq
\label{eq:superpotential}
W=\ka S(M^2-\phi^2)-\gamma S H^c\Hb^c+m\phi\pb
-\la\pb H^c\Hb^c,
\eeq
where $M$, $m$ are superheavy masses of the
order of the SUSY GUT scale $M_{\rm GUT}\simeq
2.86\times 10^{16}\units{GeV}$ and $\ka$,
$\gamma$, $\la$ are dimensionless coupling
constants. These parameters are normalized so
that they correspond to the couplings between the
SM singlet components of the superfields.
In a general superpotential of the type in
Eq.~(\ref{eq:superpotential}), $M$, $m$ and any
two of the three dimensionless parameters $\ka$,
$\gamma$, $\la$ can always be made real and
positive by appropriately redefining the
phases of the superfields. The third
dimensionless parameter, however, remains
generally complex. For definiteness, we will
choose here this parameter to be real and
positive too as we did in Ref.~\cite{nsmooth}.

\par
The F--term scalar potential obtained from the
superpotential $W$ in
Eq.~(\ref{eq:superpotential}) is given by
\bea
\label{eq:SUSYpotential}
V&=&|\ka\,(M^2-\phi^2)-\gamma H^c\Hb^c|^2
\nonumber\\
& &+|m\pb-2\ka S\phi|^2+|m\phi-\la H^c\Hb^c|^2
\nonumber\\
& &+|\gamma S+\la\pb\,|^2\left(|H^c|^2+|\Hb^c|^2
\right),
\eea
where the complex scalar fields which belong to
the SM singlet components of the superfields are
denoted by the same symbol. We will ignore
throughout the soft SUSY breaking terms
\cite{sstad} in the scalar potential since their
effect on inflationary dynamics is negligible in
our case as in the case of the conventional
realization of shifted hybrid inflation.

\par
From the potential in
Eq.~\eqref{eq:SUSYpotential} and the vanishing of
the D--terms (which implies that $\bar{H}^{c*}=
e^{i\theta}H^c$), we find \cite{nsmooth} that
there exist two distinct continua of SUSY vacua:
\bea
\phi=\phi_{+},\quad\bar{H}^{c*}=H^c,\quad |H^c|=
\sqrt{\frac{m\phi_{+}}{\la}} \quad (\theta=0);
\label{eq:vacua+}\\
\phi=\phi_{-},\quad\bar{H}^{c*}=-H^c,\quad |H^c|=
\sqrt{\frac{-m\phi_{-}}{\la}} \quad
(\theta=\pi)
\label{eq:vacua-}
\eea
with $\pb=S=0$, where
\begin{equation}
\phi_{\pm}\equiv\frac{\gamma m}{2\ka\la}\left(
-1\pm\sqrt{1+\frac{4\ka^2\la^2M^2}
{\gamma^2m^2}}\,\right).
\label{eq:SUSYvacua1}
\end{equation}
It has been shown \cite{nsmooth} that the
potential in Eq.~\eqref{eq:SUSYpotential},
generally, possesses three flat directions. The
first one is the usual trivial flat direction at
$\phi=\pb=H^c=\Hb^c=0$ with
$V=V_\text{tr}\equiv\ka^2M^4$. The second one,
which appears at
\begin{gather}
\phi=-\frac{\gamma m}{2\ka\la},\quad
\pb=-\frac{\gamma}{\la}\,S,
\nonumber \\
H^c\Hb^c=\frac{\ka\gamma(M^2-\phi^2)+\la m\phi}
{\gamma^2+\la^2},
\nonumber \\
V=V_\text{nsh}\equiv\frac{\ka^2\la^2}
{\gamma^2+\la^2}\left(M^2+\frac{\gamma^2m^2}
{4\ka^2\la^2}\right)^2,
\end{gather}
exists only for $\gamma\neq 0$ and is the
trajectory for the new shifted hybrid inflation
\cite{nshift}. Along this direction, $G_{\rm PS}$
is broken to $G_{\rm SM}$. The third flat
direction, which exists only if
$\tilde{M}^2\equiv M^2-m^2/2\ka^2>0$, lies at
\beq
\phi=\pm\,\tilde{M},\quad
\pb=\frac{2\ka\phi}{m}\,S,\quad
H^c=\Hb^c=0.
\label{eq:semishift}
\eeq
It is a ``semi-shifted'' flat direction
(in the sense that, although the field $\phi$ is
shifted from zero, the fields $H^c$, $\Hb^c$
remain zero on it) with
\beq
V=V_\text{ssh}\equiv\ka^2(M^4-\tilde{M}^4).
\eeq
Along this direction $G_{\rm PS}$ is broken to
$G_{\rm SM}\times {\rm U(1)_{B-L}}$.

\par
In our subsequent discussion, we will concentrate
on the case where $\tilde{M}^2>0$. It is
interesting to note that, in this case, the
trivial flat direction is \cite{nsmooth} unstable
as it is a path of saddle points of the
potential, while the new smooth path, which
exists for $\tilde{M}^2<0$ and was used in
Ref.~\cite{nsmooth} as
inflationary trajectory, disappears. Moreover,
for $\tilde{M}^2>0$, we
always have $V_\text{ssh}<V_\text{nsh}$. It is,
thus, more likely that the system will eventually
settle down on the semi-shifted rather than the
new shifted flat direction. Semi-shifted hybrid
inflation can then take place as the system
slowly rolls down the semi-shifted path driven by
its logarithmic slope provided by one-loop
radiative corrections \cite{rc} which are due to
the SUSY breaking by the non-vanishing potential
energy density on this path. As the system crosses
the critical point of the semi-shifted path, the
${\rm U(1)_{B-L}}$ gauge symmetry breaks
generating a network of local cosmic strings,
which contribute a small amount to the CMB
temperature power spectrum. As mentioned, for
models with local cosmic strings, it has been
shown in Ref.~\cite{bevis1} that, at $95\%$ c.l.,
$n_{\rm s}=0.94-1.06$ and $f_{10}=0.02-0.18$.

\section{One-loop Radiative Corrections}
\label{sec:radcorr}

\par
The one-loop radiative correction to the
potential on the semi-shifted path is
calculated by the Coleman-Weinberg formula
\cite{ColemanWeinberg}:
\beq
\label{eq:ColemanWeinberg}
\Delta V=
\frac{1}{64\pi^2}\;\sum_i(-1)^{F_i}M_i^4
\ln\frac{M_i^2}{\Lambda^2},
\eeq
where the sum extends over all helicity states
$i$, $F_i$ and $M_i^2$ are the fermion number and
mass squared of the $i$th state and $\Lambda$ is
a renormalization mass scale. In order to use this
formula for creating a logarithmic slope in the
inflationary potential, one has first to derive
the mass spectrum of the model on the
semi-shifted path.

\par
As mentioned, during semi-shifted hybrid
inflation, the SM singlet components of $\phi$,
$\pb$ acquire non-vanishing values and break
$G_{\rm PS}$ to
$G_{\rm SM}\times {\rm U(1)_{B-L}}$. The value
of the complex scalar field $S$ at a point of
the semi-shifted path is taken real by an
appropriate R transformation. For simplicity, we
use the same symbol $S$ for this real value of
the field as for the complex field in general
since the distinction will be obvious from the
context. The deviation of the complex scalar
field $S$ from its (real) value at a point of the
inflationary path is denoted by $\delta S$. We
can further write $\phi=v+\delta\phi$,
$\pb=\bar{v}+\delta\pb$ with $v=\pm\tilde{M}$,
$\bar{v}=(2\ka v/m)\,S$ and $\delta\phi$,
$\delta\pb$ being complex scalar fields. We can
then define the canonically normalized complex
scalar fields
\beq
\zeta=\frac{2\ka v\,\delta S-m\delta\pb}
{(m^2+4\ka^2v^2)^{1/2}},\quad
\epsilon=\frac{m\delta S+2\ka v\,\delta\pb}
{(m^2+4\ka^2v^2)^{1/2}}.
\eeq
We find that $\epsilon$ remains massless on the
semi-shifted path. So, it corresponds to the
complex scalar inflaton field
$\Sigma=(mS+2\ka v\pb)/(m^2+4\ka^2v^2)^{1/2}$,
which during inflation takes the form
$\Sigma=(1+4\ka^2v^2/m^2)^{1/2}S$. Consequently,
in our case, the real canonically normalized
inflaton is
\beq
\label{eq:inflaton}
\sigma=2^{1/2}(1+4\ka^2v^2/m^2)^{1/2}S,
\eeq
where $S$ is obviously rotated to be real.

\par
Expanding the complex scalars $\zeta$,
$\delta\phi$, $H^c$, and $\Hb^c$ in real and
imaginary parts according to the prescription
$\chi=(\chi_1+i\chi_2)/\sqrt{2}$, we find that
the mass-squared matrices $M_{-}^2$ of
$\zeta_1$, $\delta\phi_1$, $M_{+}^2$ of
$\zeta_2$, $\delta\phi_2$, $M_1^2$ of $H^c_1$,
$\Hb^c_1$, and $M_2^2$ of $H^c_2$, $\Hb^c_2$
are given by
\begin{gather}
M_{\pm}^2=m^2\left(\ba{cc}
1+a^2 & s(1+a^2)^{1/2} \\
s(1+a^2)^{1/2} & 1+a^2+s^2\pm 1
\ea\right),\\
M_{1,2}^2=m^2\left(\ba{cc}
s^2b^2 & \mp b \\
\mp b & s^2b^2
\ea\right),
\end{gather}
where $a=2\ka v/m$, $b=(\gamma+\la a)/2\ka$, and
$s=2\ka S/m$. Note that the eigenvalues of the
matrices $M_{\pm}^2$ are always positive. Though,
this is not the case with $M_{1,2}^2$.
Specifically, one of the two eigenvalues of each
of these matrices is always positive, while the
other one becomes negative for
$|s|<s_c\equiv 1/\sqrt{|b|}$ (we assume that
$b\neq 0$). This defines the critical point on
the semi-shifted path at which this path is
destabilized (see below).

\par
The superpotential in Eq.~\eqref{eq:superpotential}
gives rise to mass terms between the fermionic
partners of $\zeta$, $\delta\phi$ and $H^c$,
$\Hb^c$ (the fermionic partner of $\epsilon$
remains massless). The squares of the
corresponding mass matrices are found to be
\begin{gather}
M_0^2=m^2\left(\ba{cc}
1+a^2 & s(1+a^2)^{1/2} \\
s(1+a^2)^{1/2} & 1+a^2+s^2
\ea\right),\\
\bar{M}_0^2=m^2\left(\ba{cc}
s^2b^2 & 0 \\ 0 & s^2b^2
\ea\right).
\end{gather}

\par
This completes the analysis of the SM singlet
sector of the model. In summary, we found four
groups of two real scalars with mass-squared
matrices $M_{+}^2$, $M_{-}^2$, $M_{1}^2$, and
$M_{2}^2$ and two groups of two
Weyl fermions with mass matrices squared
$M_0^2$ and $\bar{M}_0^2$. The contribution
of the SM singlet sector to the radiative
corrections to the potential along the
semi-shifted path is given by
\bea
\label{eq:RC-SM-type}
\Delta V&=&\frac{1}{64\pi^2}\,
\Tr\Bigg\{M_{+}^4\ln\frac{M_{+}^2}{\Lambda^2}
+M_{-}^4\ln\frac{M_{-}^2}{\Lambda^2}
\nonumber \\
& &-2M_0^4\ln\frac{M_0^2}{\Lambda^2}
+M_{1}^4\ln\frac{M_{1}^2}{\Lambda^2}
+M_{2}^4\ln\frac{M_{2}^2}{\Lambda^2}
\nonumber \\
& &-2\bar{M}_0^4\ln\frac{\bar{M}_0^2}
{\Lambda^2}\Bigg\}.
\eea

\par
We now turn to the $u^c$, $\bar{u}^c$ type fields
which are color antitriplets with charge $-2/3$
and color triplets with charge $2/3$
respectively. Such fields exist in $H^c$,
$\Hb^c$, $\phi$, and $\pb$ and we shall denote
them by $u_H^c$, $\bar{u}_H^c$, $u_\phi^c$,
$\bar{u}_\phi^c$, $u_{\pb}^c$, and
$\bar{u}_{\pb}^c$. The relevant expansion of
$\phi$ is
\bea
\label{eq:phi-u-exp}
\phi&=&\left[\frac{1}{\sqrt{12}}\left(
\ba{cc}\bm{1}_3 & 0\\0 & -3\ea\right)
\; ,\;\frac{1}{\sqrt{2}}\left(
\ba{cc}1 & 0\\0 & -1\ea\right)
\right]\phi\nonumber\\
& &+
\left(\ba{cc}0 & \bm{0}_3\\1 & 0\ea\right)
u_\phi^c+
\left(\ba{cc}0 & 1\\\bm{0}_3 & 0\ea\right)
\bar{u}_\phi^c+\dots,
\eea
where the SM singlet in $\phi$ (denoted by the
same symbol) is also shown with the first
(second) matrix in the brackets belonging to the
algebra of ${\rm SU}(4)_c$ (${\rm SU(2)_R}$).
Here, $\bm{1}_3$ and $\bm{0}_3$ denote the
$3\times 3$ unit and zero matrices respectively.
The fields $u_\phi^c$, $\bar{u}_\phi^c$ are
${\rm SU(2)_R}$ singlets, so only their
${\rm SU}(4)_c$ structure is shown and summation
over their ${\rm SU}(3)_c$ indices is implied in
the ellipsis. The field $\pb$ can be similarly
expanded.

\par
In the bosonic $u^c$, $\bar{u}^c$ type sector,
we find that the mass-squared matrices
$M_{u\pm}^2$ of the complex scalar fields
$u_{\chi\pm}^c=(u_{\chi}^c
\pm\bar{u}_{\chi}^{c*})/\sqrt{2}$, for
$\chi=H,\phi,\pb$, are
\begin{gather}
\label{eq:Mu+}
M_{u+}^2=m^2\left(\ba{ccc}
c^2s^2-c & 0 & 0\\
0 & s^2 & -s\\
0 & -s & 1\ea\right),\\
\label{eq:Mu-}
M_{u-}^2=m^2\left(\ba{ccc}
c^2s^2+c & 0 & 0\\ [2pt]
0 & 2+s^2+\rho_g^2 & -s(1-\rho_g^2)\\ [3pt]
0 & -s(1-\rho_g^2) & 1+\rho_g^2s^2
\ea\right),
\end{gather}
where $c=(\gamma-\la a/3)/2\ka$ and
$\rho_g^2=g^2a^2/3\ka^2$ with $g$ being the
$G_{\rm PS}$ gauge coupling constant. Note that
$\rho_g^2$ parametrizes contributions arising
from the D--terms of the scalar potential and
$M_{u+}^2$ has one zero eigenvalue corresponding
to the Goldstone boson which is absorbed by the
superhiggs mechanism. Furthermore, one of the
eigenvalues $m^2(c^2s^2\mp c)$ of the matrices in
Eqs.~(\ref{eq:Mu+}) and (\ref{eq:Mu-})
(depending on the sign of $c$) becomes negative
as soon as $s$ crosses below the point $s_c^{(1)}
\equiv 1/\sqrt{|c|}$ on the semi-shifted path.
So, if $s_c^{(1)}$ was larger than the critical
value $s_c$, the system would be
destabilized first in one of the directions
$u_{H\pm}^c$. In this case, a
${\rm SU}(3)_c$-breaking VEV would
develop. To avoid this, we should demand that
$s_c^{(1)}$ is located lower than the critical
point $s_c$, so that, after the end of inflation,
the correct symmetry breaking is obtained. This
gives the condition $|b|<|c|$, which we will
consider later.

\par
In the fermionic $u^c$, $\bar{u}^c$ type sector,
we obtain four Dirac fermions (per color):
$\psi_{u_H^c}^D=\psi_{u_H^c}+\psi^c_{\bar{u}_H^c}$,
$\psi_{u_{\phi}^c}^D=\psi_{u_{\phi}^c}+
\psi^c_{\bar{u}_{\phi}^c}$, $\psi_{u_{\pb}^c}^D=
\psi_{u_{\pb}^c}+\psi^c_{\bar{u}_{\pb}^c}$, and
$-i\la^D=-i(\la^+ + \la^{-c})$. Here, $\psi_\chi$
is the fermionic partner of the complex scalar
field $\chi$ and
$\la^{\pm}=(\la^1\pm i\la^2)/\sqrt{2}$, where
$\la^1$ ($\la^2$) is the gaugino color triplet
corresponding to the ${\rm SU}(4)_c$ generators
with $1/2$ ($-i/2$) in the $i4$ and $1/2$ ($i/2$)
in the $4i$ entry ($i=1,2,3$). The fermionic
mass matrix is
\beq
M_{\psi_u}=m\left(\ba{cccc}
-cs & 0 & 0 & 0\\
0 & -s & 1 & -\rho_g\\
0 & 1 & 0 & -\rho_g s\\
0 & -\rho_g & -\rho_g s & 0
\ea\right).
\eeq
To complete this sector, we must also include the
gauge bosons $A^{1,2}$ which are associated with
$\la^{1,2}$. They acquire a mass squared
$M_g^2=m^2\rho_g^2(1+s^2)$.

\par
The overall contribution of the $u^c$, $\bar{u}^c$
type sector to $\Delta V$ in
Eq.~\eqref{eq:ColemanWeinberg} is
\bea
\label{eq:RC-u-type}
\Delta V&=&\frac{3}{32\pi^2}
\Tr\Bigg\{M_{u+}^4\ln\frac{M_{u+}^2}
{\Lambda^2}
+M_{u-}^4\ln\frac{M_{u-}^2}{\Lambda^2}
\nonumber \\
& &-2M_{\psi_u}^4\ln\frac{M_{\psi_u}^2}{\Lambda^2}
+3M_g^4\ln\frac{M_g^2}{\Lambda^2}\Bigg\}.
\eea

\par
We will now discuss the contribution from the
$e^c$, $\bar{e}^c$ type sector consisting of
color singlets with charge $1$, $-1$. Such
fields exist in $H^c$, $\Hb^c$, $\phi$, and
$\pb$ and we shall denote them by $e_H^c$,
$\bar{e}_H^c$, $e_\phi^c$, $\bar{e}_\phi^c$,
$e_{\pb}^c$, and $\bar{e}_{\pb}^c$. The relevant
expansion of $\phi$ is
\beq\label{eq:phi-e-exp}
\phi=\left[\frac{1}{\sqrt{12}}\left(
\ba{cc}\bm{1}_3 & 0\\0 & -3\ea\right)
\; ,\;\left(
\ba{cc}0 & 1\\0 & 0\ea\right)e_\phi^c
+\left(
\ba{cc}0 & 0\\1 & 0\ea\right)\bar{e}_\phi^c
\right]
\eeq
with the same notation as in
Eq.~\eqref{eq:phi-u-exp}. A similar expansion
holds for $\pb$. It turns out that the mass terms
in this sector are exactly the same as in the
$u^c$, $\bar{u}^c$ type sector with $\lambda/3$
replaced by $\lambda$ and $2g^2/3$ by $g^2$. So,
we will only summarize the results.

\par
In the bosonic $e^c$, $\bar{e}^c$ type sector,
the mass-squared matrices $M_{e\pm}^2$ of the
complex scalars $e_{\chi\pm}^c=(e_{\chi}^c\pm
\bar{e}_{\chi}^{c*})/\sqrt{2}$, for
$\chi=H,\phi,\pb$, are
\begin{gather}
\label{eq:Me+}
M_{e+}^2=m^2\left(\ba{ccc}
d^2s^2-d & 0 & 0\\
0 & s^2 & -s\\
0 & -s & 1\ea\right),\\
\label{eq:Me-}
M_{e-}^2=m^2\left(\ba{ccc}
d^2s^2+d & 0 & 0\\ [2pt]
0 & 2+s^2+\tau_g^2 & -s(1-\tau_g^2)\\ [3pt]
0 & -s(1-\tau_g^2) & 1+\tau_g^2s^2
\ea\right),
\end{gather}
where $d=(\gamma-\la a)/2\ka$ and $\tau_g=
\sqrt{3/2}\,\rho_g$. Note that, again,
$M_{e+}^2$ has one zero eigenvalue
corresponding to the Goldstone boson which
is absorbed by the superhiggs mechanism.
Furthermore, one of the eigenvalues
$m^2(d^2s^2\mp d)$ of the matrices in
Eqs.~(\ref{eq:Me+}) and(\ref{eq:Me-}) (depending
on the sign of $d$) becomes negative as $s$
crosses below $s_c^{(2)}\equiv 1/\sqrt{|d|}$ on
the semi-shifted path. Therefore, we must impose
the constraint $s_c^{(2)}<s_c\Rightarrow
|b|<|d|$ for the same reason explained above.

\par
In the fermionic $e^c$, $\bar{e}^c$ type sector,
we obtain four Dirac fermions with mass matrix
\beq
M_{\psi_e}=m\left(\ba{cccc}
-ds & 0 & 0 & 0\\
0 & -s & 1 & -\tau_g\\
0 & 1 & 0 & -\tau_g s\\
0 & -\tau_g & -\tau_g s & 0
\ea\right).
\eeq
Finally, we again obtain two gauge bosons with
mass squared $\hat{M}_g^2=m^2\tau_g^2(1+s^2)$.

\par
The overall contribution of the $e^c$,
$\bar{e}^c$ type sector to $\Delta V$ in
Eq.~\eqref{eq:ColemanWeinberg} is
\bea
\label{eq:RC-e-type}
\Delta V&=&\frac{1}{32\pi^2}
\Tr\Bigg\{M_{e+}^4\ln\frac{M_{e+}^2}
{\Lambda^2}
+M_{e-}^4\ln\frac{M_{e-}^2}{\Lambda^2}
\nonumber \\
& &-2M_{\psi_e}^4\ln\frac{M_{\psi_e}^2}{\Lambda^2}
+3\hat{M}_g^4\ln\frac{\hat{M}_g^2}{\Lambda^2}
\Bigg\}.
\eea

\par
Let us now consider the $d^c$, $\bar{d}^c$ type
sector consisting of color antitriplets with
charge $1/3$ and color triplets with charge
$-1/3$. Such fields exist in $H^c$, $\Hb^c$,
$\phi$, and $\pb$ and we denote them by $d_H^c$,
$\bar{d}_H^c$, $d_\phi^c$, $\bar{d}_\phi^c$,
$d_{\pb}^c$, and $\bar{d}_{\pb}^c$. The field
$\phi$ can be expanded in terms of these fields
as
\bea\label{eq:phi-d-exp}
\phi&=&\left[\left(
\ba{cc}0 & \bm{0}_3\\1 & 0\ea\right)
\; ,\;\left(
\ba{cc}0 & 1\\0 & 0\ea\right)\right]d_\phi^c
\nonumber\\
& &+\left[\left(
\ba{cc}0 & 1\\\bm{0}_3 & 0\ea\right)
\; ,\;\left(
\ba{cc}0 & 0\\1 & 0\ea\right)\right]
\bar{d}_\phi^c+\dots
\eea
with the notation of Eq.~\eqref{eq:phi-u-exp}.
The field $\pb$ is similarly expanded.

\par
In the bosonic $d^c$, $\bar{d}^c$ type sector,
the mass-squared matrices $M_{d\pm}^2$ of the
complex scalars $d_{\chi\pm}^c=(d_{\chi}^c\pm
\bar{d}_{\chi}^{c*})/\sqrt{2}$, for
$\chi=H,\phi,\pb$, are
\beq
M_{d\pm}^2=m^2\left(\ba{ccc}
e^2s^2\mp e & 0 & 0\\
0 & 1+s^2\mp 1 & -s\\
0 & -s & 1\ea\right),
\eeq
where $e=(\gamma+\la a/3)/2\ka$. Note that,
again, one of the eigenvalues $m^2(e^2s^2\mp e)$
of these matrices (depending on the sign of $e$)
becomes negative as $s$ crosses below
$s_c^{(3)}\equiv 1/\sqrt{|e|}$ on the
semi-shifted path and we, thus, have to impose
the constraint $s_c^{(3)}<s_c\Rightarrow
|b|<|e|$, so that the correct symmetry breaking
pattern occurs at the end of inflation.

\par
In the fermionic $d^c$, $\bar{d}^c$ type sector,
we obtain three Dirac fermions (per color) with
mass matrix
\beq
M_{\psi_d}=m\left(\ba{ccc}
-es & 0 & 0\\
0 & -s & 1\\
0 & 1 & 0
\ea\right).
\eeq
Note that there are no D--terms, gauge bosons, or
gauginos in this sector.

\par
The contribution of this sector to $\Delta V$
in Eq.~\eqref{eq:ColemanWeinberg} is
\bea
\label{eq:RC-d-type}
\Delta V&=&\frac{3}{32\pi^2}
\Tr\Bigg\{M_{d+}^4\ln\frac{M_{d+}^2}
{\Lambda^2}
+M_{d-}^4\ln\frac{M_{d-}^2}{\Lambda^2}
\nonumber \\
& &-2M_{\psi_d}^4\ln\frac{M_{\psi_d}^2}{\Lambda^2}
\Bigg\}.
\eea

\par
Next, we consider the $q^c$, $\bar{q}^c$
type fields which are color antitriplets with
charge $-5/3$ and color triplets with charge
$5/3$. They exist in $\phi$, $\pb$ and we call
them $q_\phi^c$, $\bar{q}_\phi^c$, $q_{\pb}^c$,
$\bar{q}_{\pb}^c$. The relevant expansion of
$\phi$ is
\bea\label{eq:phi-q-exp}
\phi&=&\left[\left(
\ba{cc}0 & \bm{0}_3\\1 & 0\ea\right)
\; ,\;\left(
\ba{cc}0 & 0\\1 & 0\ea\right)\right]q_\phi^c
\nonumber\\
& &+\left[\left(
\ba{cc}0 & 1\\\bm{0}_3 & 0\ea\right)
\; ,\;\left(
\ba{cc}0 & 1\\0 & 0\ea\right)\right]
\bar{q}_\phi^c+\dots
\eea
and a similar expansion holds for $\pb$.

\par
In the bosonic $q^c$, $\bar{q}^c$ type sector,
the mass-squared matrices $M_{q\pm}^2$ of the
complex scalars $q_{\chi\pm}^c
=(q_{\chi}^c\pm\bar{q}_{\chi}^{c*})/\sqrt{2}$, for
$\chi=\phi,\pb$, are
\beq
M_{q\pm}^2=m^2\left(\ba{cc}
1+s^2\mp 1 & -s\\
-s & 1\ea\right).
\eeq

\par
In the fermionic $q^c$, $\bar{q}^c$ type sector,
we obtain two Dirac fermions (per color) with
mass matrix
\beq
M_{\psi_q}=m\left(\ba{cc}
-s & 1\\1 & 0\ea\right).
\eeq
There are no D--terms, gauge bosons, or
gauginos in this sector as well.

\par
The contribution of this sector to $\Delta V$
in Eq.~\eqref{eq:ColemanWeinberg} is
\bea
\label{eq:RC-q-type}
\Delta V&=&\frac{3}{32\pi^2}
\Tr\Bigg\{M_{q+}^4\ln\frac{M_{q+}^2}{\Lambda^2}
+M_{q-}^4\ln\frac{M_{q-}^2}{\Lambda^2}
\nonumber \\
& &-2M_{\psi_q}^4\ln\frac{M_{\psi_q}^2}{\Lambda^2}
\Bigg\}.
\eea

\par
Finally, in $\phi$, $\pb$ there exist color octet,
$\rm{SU(2)_R}$ triplet superfields: $\phi_8^0$,
$\phi_8^{\pm}$, $\pb_8^0$, $\pb_8^{\pm}$, with
charge $0$, $1$, $-1$ as indicated. The relevant
expansion of $\phi$ is
\bea\label{eq:phi-p8-exp}
\phi&=&\left[\left(
\ba{cc}T_8 & 0\\0 & 0\ea\right)
\; ,\;\frac{1}{\sqrt{2}}\left(
\ba{cc}1 & 0\\0 & -1\ea\right)\phi_8^0
\right.\nonumber\\
& &+\left.\left(
\ba{cc}0 & 1\\0 & 0\ea\right)\phi_8^+ +
\left(\ba{cc}0 & 0\\1 & 0\ea\right)\phi_8^-
\right]+\dots,
\eea
where $T_8$ represents the eight ${\rm SU}(3)_c$
generators appropriately normalized. A similar
expansion holds for $\pb$.

\par
In the bosonic sector, we obtain two groups of
24 complex scalars, which can be combined in
pairs of two with mass-squared matrix
\beq
M_{\phi_8\pm}^2=m^2\left(\ba{cc}
1+s^2\mp 1 & -s\\
-s & 1\ea\right).
\eeq

\par
In the fermionic sector, we find 48 Weyl fermions
which can be combined in pairs of two with
mass matrix
\beq
M_{\psi_{\phi_8}}=m\left(\ba{cc}
-s & 1\\
1 & 0\ea\right).
\eeq

\par
The contribution of this sector to $\Delta V$
in Eq.~\eqref{eq:ColemanWeinberg} is
\bea
\label{eq:RC-p8-type}
\Delta V&=&\frac{12}{32\pi^2}
\Tr\Bigg\{M_{\phi_8+}^4
\ln\frac{M_{\phi_8+}^2}{\Lambda^2}
\nonumber \\
& &+M_{\phi_8-}^4
\ln\frac{M_{\phi_8-}^2}{\Lambda^2}
-2M_{\psi_{\phi_8}}^4\ln\frac{M_{\psi_{\phi_8}}^2}
{\Lambda^2}\Bigg\}. \nonumber \\
\eea

\par
The final overall $\Delta V$ is found by adding
the contributions from the SM singlet sector in
Eq.~\eqref{eq:RC-SM-type}, the $u^c$, $\bar{u}^c$
type sector in Eq.~\eqref{eq:RC-u-type}, the
$e^c$, $\bar{e}^c$ type sector in
Eq.~\eqref{eq:RC-e-type}, the $d^c$, $\bar{d}^c$
type sector in Eq.~\eqref{eq:RC-d-type}, the
$q^c$, $\bar{q}^c$ type sector in
Eq.~\eqref{eq:RC-q-type}, and the color octet
sector in Eq.~\eqref{eq:RC-p8-type}. These
one-loop radiative corrections are added to the
tree-level potential $V_{\rm ssh}$ yielding the
effective potential along the semi-shifted
inflationary path in global SUSY.
They generate a slope on this path which is
necessary for driving the system toward the
vacuum. The overall $\sum_i(-1)^{F_i}M_i^4$ is
$\si$-independent, which implies that the
overall slope of the effective potential is
$\Lambda$-independent. This is a crucial
property of the model since otherwise
observable quantities like the power spectrum
$P_{\mathcal R}$ of the primordial curvature
perturbation or the spectral index would
depend on the scale $\Lambda$, which remains
undetermined.

\par
Let us now discuss the constraints
$0<|b|<|c|,|d|,|e|$ derived in the course of
the calculation of the mass spectrum on the
semi-shifted path. It is easy to show that
these constraints require that $v$ be in
one of the ranges
\beq
\label{eq:constraints}
0>v>-\frac{\gamma m}{2\ka\la}
\quad\text{or}\quad
-\frac{\gamma m}{2\ka\la}>v>
-\frac{3\gamma m}{4\ka\la}.
\eeq
These two ranges of $v$ lead, respectively, to
the two different sets of SUSY vacua of
Eqs.~\eqref{eq:vacua+} and \eqref{eq:vacua-}. To
see this, let us replace all the fields in the
scalar potential of Eq.~\eqref{eq:SUSYpotential}
except $H^c$, $\bar{H}^c$ by their values on the
semi-shifted path. Taking into account that
$\bar{H}^{c*}=e^{i\theta}H^c$ from the vanishing
of the D--terms, we are then left with only two
free degrees of freedom, namely $|H^c|$ and
$\theta$, and the potential becomes
\beq
V=V_{\rm ssh}+2m^2b^2\left(
s^2-\frac{\cos\theta}{b}\right)|H^c|^2
+(\gamma^2+\la^2)|H^c|^4.
\eeq
It is obvious from this equation that, if $b>0$,
which is the case in the first range for $v$ in
Eq.~\eqref{eq:constraints}, the system will
get destabilized toward the direction with
$\cos\theta=1$ leading to the SUSY vacua
in Eq.~\eqref{eq:vacua+}, while, if $b<0$, which
holds in the second range for $v$ in
Eq.~\eqref{eq:constraints}, the system will be
led to the SUSY vacua in Eq.~\eqref{eq:vacua-}.

\section{Supergravity corrections}
\label{sec:sugra}

\par
We now turn to the discussion of the SUGRA
corrections to the inflationary potential of the
model. The F--term scalar potential in SUGRA is
given by
\beq
\label{eq:VSUGRA}
V=e^{K/\mP^2}
\left[(F_i)^* K^{i^*j}F_j-3\,\frac{|W|^2}
{\mP^2}\right],
\eeq
where $K$ is the K\"{a}hler potential, $\mP$ the
reduced Planck mass, $F_i=W_i+K_iW/\mP^2$, a
subscript $i$ ($i^*$) denotes derivation with
respect to the complex scalar field $\chi^i$
($\chi^{i{\,*}}$), and $K^{i^*j}$ is the inverse
of the K\"{a}hler metric $K_{j\,i^*}$. We will
consider SUGRA with minimal K\"{a}hler potential
and show that the results of the fit in
Ref.~\cite{bevis1} can be naturally met.

\par
The minimal K\"{a}hler potential in the model
under consideration has the form
\beq
\label{eq:MinKahler}
K^{\rm min}=|S|^2+|\phi|^2+|\pb|^2+|H^c|^2+
|\Hb^c|^2
\eeq
and the corresponding F--term scalar potential is
\beq
\label{eq:SUGRAmin}
V^{\rm min}=e^{K^{\rm min}/\mP^2}\;\left[
\sum_{\chi}\left|W_\chi+\frac{W\chi^*}{\mP^2}
\right|^2-3\,\frac{|W|^2}{\mP^2}\right],
\eeq
where $\chi$ stands for any of the five complex
scalar fields appearing in
Eq.~\eqref{eq:MinKahler}. It is quite easily
verified that, on the semi-shifted direction,
this scalar potential expanded up to fourth
order in $|S|$ takes the form (the SUGRA
corrections to the location of the semi-shifted
path are not taken into account since they are
small)
\beq
V^{\rm min}_{\rm ssh}\simeq V_{\rm ssh}\,
e^{\frac{\tilde{M}^2}{\mP^2}}\left[
1+\frac{1}{2}
\frac{\tilde{M}^2}{\mP^2}\frac{\si^2}{\mP^2}+
\frac{1}{8}\left(1+\frac{2\tilde{M}^2}{\mP^2}
\right)\frac{\si^4}{\mP^4}\right],
\eeq
where $V_{\rm ssh}$ is the constant classical
energy density on the semi-shifted path in the
global SUSY case and $\si$ is the canonically
normalized inflaton field defined in
Eq.~\eqref{eq:inflaton}. Thus, after including
the SUGRA corrections with minimal K\"{a}hler
potential, the effective potential during
semi-shifted hybrid inflation becomes
\beq
\label{Vtrsugra}
V^{\rm mSUGRA}_{\rm ssh}\simeq
V^{\rm min}_{\rm ssh}+\Delta V
\eeq
with $\Delta V$ representing the overall
one-loop radiative correction calculated in
Sec.~\ref{sec:radcorr}.

\section{Inflationary Observables}
\label{sec:observables}

\par
The slow-roll parameters $\veps$, $\eta$
and the parameter $\xi^2$, which enters the
running of the spectral index, are given
(see e.g. Ref.~\cite{review}) by
\bea
\veps &\equiv& \frac{\mP^2}{2}\,
\left(\frac{V^\prime(\si)}{V(\si)}\right)^2,\\
\eta &\equiv& \mP^2\,\left(
\frac{V^{\prime\prime}(\si)}{V(\si)}\right),\\
\xi^2 &\equiv& \mP^4
\left(\frac{V^\prime(\si)V^{\prime\prime\prime}
(\si)}{V^2(\si)}\right),
\eea
where prime denotes derivation with respect to
the real canonically normalized inflaton field
$\si$ defined in Eq.~\eqref{eq:inflaton}. Here
and in the subsequent formulas in
Eqs.~(\ref{eq:nq}) and (\ref{eq:Perturbations}),
$V$ is the effective potential
$V^{\rm mSUGRA}_{\rm ssh}$ defined in
Eq.~(\ref{Vtrsugra}). Inflation ends at
$\si_f=\max\{\si_\eta,\si_c\}$, where
$\si_\eta>0$ denotes the value of the inflaton
field when $\eta=-1$ and $\si_c>0$ is the
critical value of $\si$ on the semi-shifted
inflationary path corresponding to $s_c$.

\par
The number of e-foldings from the time when the
pivot scale $k_0=0.002~{\rm Mpc}^{-1}$ crosses
outside the inflationary horizon until the end of
inflation is (see e.g. Ref.~\cite{review})
\beq
\label{eq:nq}
N_Q\simeq\frac{1}{\mP^2}\,
\int_{\si_f}^{\si_Q}
\frac{V(\si)}{V^{\prime}(\si)}\,d\si,
\eeq
where $\si_Q$ is the value of the inflaton field
at horizon crossing of the scale $k_0$. The
inflation power spectrum $P_{\mathcal R}$ of the
primordial curvature perturbation at the pivot
scale $k_0$ is given (see e.g.
Ref.~\cite{review}) by
\beq\label{eq:Perturbations}
P_{\mathcal R}^{1/2}\simeq
\frac{1}{2\pi\sqrt{3}}\,\frac{V^{3/2}(\si_Q)}
{\mP^3V^{\prime}(\si_Q)}.
\eeq
The spectral index $n_{\rm s}$, the
tensor-to-scalar ratio $r$, and the running of
the spectral index $dn_{\rm s}/d\ln k$ are
written (see e.g. Ref.~\cite{review}) as
\begin{gather}
n_{\rm s}\simeq 1+2\eta-6\veps,\quad
r\simeq\,16\veps,\nonumber\\
\frac{dn_{\rm s}}{d\ln k}\simeq16\veps\eta
-24\veps^2-2\xi^2,\nonumber
\end{gather}
where $\veps$, $\eta$, and $\xi^2$ are evaluated
at $\si=\si_Q$. The number of e-foldings $N_Q$
required for solving the horizon and flatness
problems of standard hot big bang cosmology is
approximately given (see e.g.
Ref.~\cite{lectures}) by
\beq\label{eq:NQvsVinf}
N_Q\simeq53.76\,+\frac{2}{3}\,\ln\left(\frac{v_0}
{10^{15}\units{GeV}}\right)+\frac{1}{3}\,\ln
\left(\frac{T_{\rm r}}{10^9\units{GeV}}\right),
\eeq
where $v_0=V_{\rm ssh}^{1/4}$ is the inflationary
scale and $T_{\rm r}$ is the reheat temperature
that is expected not to exceed about
$10^9\units{GeV}$, which is the well-known
gravitino bound \cite{gravitino}. In the
following, we take $T_{\rm r}$ to saturate the
gravitino bound, i.e.
$T_{\rm r}=10^9\units{GeV}$.

\section{String Power Spectrum}
\label{sec:string}

\par
As mentioned before, the spontaneous breaking of
the ${\rm U(1)_{B-L}}$ gauge symmetry at the end
of the semi-shifted hybrid inflation leads to the
formation of local cosmic strings. These strings
can contribute a small amount to the CMB power
spectrum. Their contribution is parametrized
\cite{bevis1} to a very good approximation by the
dimensionless string tension $G\mu_{\rm s}$,
where $G$ is the Newton's gravitational constant
and $\mu_{\rm s}$ is the string tension, i.e. the
energy per unit length of the string. In
Refs.~\cite{bevis1,bevis2}, local strings were
considered within the Abelian Higgs model in the
Bogomol'nyi limit, i.e. with equal scalar and
vector particle masses. If this was the case in
our model, the string tension would be given by
\beq
\label{eq:stringtension}
\mu_{\rm s}=4\pi|\vev{H^c}|^2,
\eeq
where $\vev{H^c}$ is the VEV of $H^c$ in the
relevant SUSY vacuum and is responsible for the
spontaneous breaking of the ${\rm U(1)_{B-L}}$
gauge symmetry. However, as it turns out, the
scalar-to-vector mass ratio in our model is
somewhat smaller than unity. This is, though, not
expected \cite{private} to make any appreciable
qualitative difference. Also, the strings in our
model do not coincide with the strings in the
simple Abelian Higgs model due to the presence of
the field $\phi$, which enters the string
solution. We do not anticipate, however, that
this will alter the picture in any essential way.
Moreover, as one can show by using the results of
Ref.~\cite{superconduct}, charged fermionic
transverse zero energy modes do not exist in the
presence of our strings, which, thus, do not
exhibit fermionic superconductivity. Therefore,
we will apply the results of
Refs.~\cite{bevis1,bevis2} in our model and adopt
the formula in Eq.~(\ref{eq:stringtension}) for
the string tension. This is certainly an
approximation, but we believe that it is adequate
for our purposes here. In Ref.~\cite{bevis1}, it
was found that the best-fit value of the string
tension required to normalize the WMAP
temperature power spectrum at multipole $\ell=10$
is
\beq
\label{eq:Gmu}
G\mu_{\rm s}=2.04\times 10^{-6}.
\eeq
This corresponds to $f_{10}=1$, which is, of
course, unrealistically large. The actual value
of $f_{10}$ is proportional to the actual value
of $(G\mu_{\rm s})^2$. So, for any given value of
$f_{10}$, we can calculate $\mu_{\rm s}$ using
its normalization in Eq.~(\ref{eq:Gmu}). From
Eq.~(\ref{eq:stringtension}), we can then
determine $|\vev{H^c}|$.

\begin{figure}[tp]
\centering
\includegraphics[width=\linewidth]{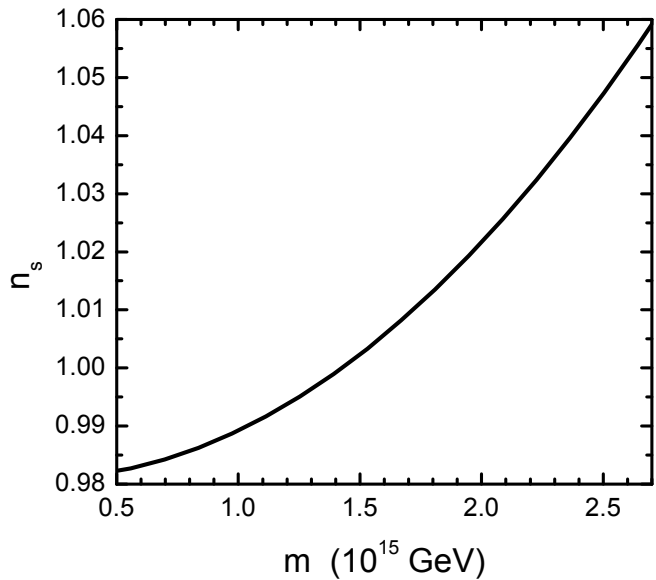}
\caption{Spectral index in semi-shifted hybrid
inflation as a function of the mass parameter $m$
in minimal SUGRA for
$v=-\gamma m/4\kappa\lambda$,
$\gamma/2\lambda=1$, and $f_{10}=0.10$.}
\label{fig:spectral}
\end{figure}

\section{Numerical Results}
\label{sec:numeric}

\par
We choose the value $v$ of the field $\phi$ on
the semi-shifted path to lie in the first range
for $v$ in Eq.~\eqref{eq:constraints}. In
particular, we take it to be in the middle of
this range, i.e.
\beq
v=-\frac{\gamma m}{4\ka\la}.
\eeq
This means, as we explained, that the universe
will end up in the vacuum of
Eq.~\eqref{eq:vacua+}. Similar results can be
obtained if one chooses the value of $v$ to be
in the second range of
Eq.~\eqref{eq:constraints}. In order to fully
determine the five parameters of the model,
we need to make another four choices. One of
them is taken to be the ratio $\gamma/2\la=1$.
Later we will comment on the dependence of the
results on variations of this ratio, which is
anyway weak. Secondly, we require the
inflationary power spectrum amplitude of the
primordial curvature perturbation at the pivot
scale $k_0$ to have its central value
\cite{private} in the fit of Ref.~\cite{bevis1}:
\beq
P_{\mathcal R}^{1/2}\simeq 4.47\times 10^{-5}.
\label{eq:PRvalue}
\eeq
Further, we take, as an example, $f_{10}$ to be
equal to 0.10, its central value \cite{bevis1}.
This determines $|\vev{H^c}|$ as discussed in
Sec.~\ref{sec:string}. Finally, we calculate the
spectral index for various values of the mass
parameter $m$. The results are presented in
Fig.~\ref{fig:spectral}, where $m$ is restricted
to be below $2.7\times 10^{15}\units{GeV}$, so
that the spectral index remains within its
$95\%$ c.l. range.

\par
For $m$ varying in the interval
$(0.5-2.7)\times 10^{15}\units{GeV}$, which is
depicted in Fig.~\ref{fig:spectral}, the ranges
of the various parameters of the model are:
$M\simeq(0.6-3.5)\times 10^{15}\units{GeV}$,
$\gamma\simeq 0.029-0.914$,
$\lambda\simeq 0.0145-0.457$,
$\ka\simeq 0.73-0.67$,
$\si_Q\simeq(0.4-3.3)\times 10^{17}\units{GeV}$,
$\si_f\simeq(1.8-5.3)\times 10^{16}\units{GeV}$,
$N_Q\simeq 53.2-54.4$,
$dn_{\rm s}/d\ln k\simeq-(0.1-3.1)\times
10^{-6}$, $r\simeq(0.001-4.5)\times 10^{-5}$, and
the ratio $\si_f/\si_c\simeq 2.6-7.7$. As one
observes, we easily achieve spectral indices that
are compatible with the fit of
Ref.~\cite{bevis1}. In particular, the best-fit
value of the spectral index $n_{\rm s}$ ($=1.00$)
is achieved for
$m\simeq 1.40\times 10^{15}\units{GeV}$. However,
indices lower than about 0.98 are not obtainable.
Actually, as we lower $m$, the SUGRA corrections
become less and less important and the spectral
index decreases tending to its value
($\approx 0.98$) in global SUSY. In all cases,
both the running of the spectral index and the
tensor-to-scalar ratio are negligibly small.

\par
Note that our results turn out to be quite
sensitive to small changes of $\la$ (and, thus,
$\gamma$). This is due to the fact that the
radiative correction to the inflationary
potential contains logarithms with large positive
as well as logarithms with large negative
inclination with respect to $\si$. If no
cancellation is assumed between these two
competing trends, one ends up with either a
rather fast rolling of the inflaton (dominance of
logarithms with large positive inclination) or a
negative inclination of the effective potential
for large values of $\si$ (dominance of
logarithms with large negative inclination). In
the latter case, after the inclusion of minimal
SUGRA corrections, which lift the potential for
$\si\gsim\mP$, a local minimum and maximum will
be generated on the inflationary path. This leads
\cite{king,rehman} to complications and must,
therefore, be avoided. It turns out that a
cancellation to the third significant digit
between the positive and negative contributions
to the derivative of the effective potential is
needed in order to avoid these complications and
ensure that the slow-roll conditions for the
inflaton are fulfilled. This can be achieved by a
mild tuning of the parameter $\la$ to the third
significant digit. So, the model entails a
moderate tuning in one of its parameters in order
to be cosmologically viable. Note, however, that
this tuning needs only to be performed between
the various contributions to the radiative
correction and it is not spoiled by minimal SUGRA
corrections. We should also mention that, in our
model, $\si_f$ turns out to be much larger than
$\si_c$ and inflation terminates well before the
system reaches the critical point of the
semi-shifted path. This is again due to the
presence in the inflationary potential of
logarithms with large inclination. Finally, we
find that reducing the ratio $\gamma/2\la$
generally leads to a slight increase of the
spectral index. Though, this dependence is rather
weak and that is why we have chosen to constrain
this ratio to a constant value (instead of
setting e.g. the ratio $\ka/\la={\rm const.}$).

\par
We observe numerically that, varying $f_{10}$
within its $95\%$ c.l. range $0.02-0.18$, the
value of $n_{\rm s}$ changes only in the third
decimal place. So, the curve in
Fig.~\ref{fig:spectral} is practically
independent of $f_{10}$. We should, however,
keep in mind that, for large values of $m$ and
low $f_{10}$'s, the constraint in
Eq.~(\ref{eq:PRvalue}) cannot be satisfied.
Consequently, the curve in
Fig.~\ref{fig:spectral} applied to low values
of $f_{10}$ terminates on the right at a value
of $m$ which, of course, depends on $f_{10}$,
but is, in any case, higher than about
$2\times 10^{15}\units{GeV}$.

\par
We have seen that, in minimal SUGRA, the model
develops a preference for values of $m$ near
$1.4\times 10^{15}\units{GeV}$. On the other
hand, for $f_{10}=0.10$, the prediction for the
value of $m$ which is derived from gauge coupling
constant unification is
$m\simeq 2.085\times 10^{15}\units{GeV}$, as the
reader may find out in Sec.~\ref{sec:gauge}.
However, one can see that, for this value of $m$,
the predicted spectral index is
$n_{\rm s}\simeq 1.0254$, which lies inside the
$1-\si$ range for $n_{\rm s}$ given by the fit
that we have been using in this paper.

\section{Gauge unification}
\label{sec:gauge}

\par
We will now discuss the question of gauge
coupling constant unification in our model. As
already mentioned, the VEVs of the fields $H^c$,
$\bar{H}^c$ break the PS gauge group $G_{\rm PS}$
to $G_{\rm SM}$, whereas the VEV of the field
$\phi$ breaks it only to
$G_{\rm SM}\times{\rm U(1)_{B-L}}$. So, the gauge
boson $A^\perp$ corresponding to the linear
combination of ${\rm U(1)_Y}$ and
${\rm U(1)_{B-L}}$ which is perpendicular to
${\rm U(1)_Y}$ acquires its mass squared
$m^2_{A^\perp}=(5/2)g^2|\vev{H^c}|^2$ solely from
the VEVs of $H^c$, $\bar{H}^c$. On the other
hand, the masses squared $m_A^2$ and
$m_{W_{\rm R}}^2$ of the color triplet,
antitriplet ($A^\pm$) and charged
${\rm SU(2)_R}$ ($W^\pm_{\rm R}$)
gauge bosons get contributions from $\vev{\phi}$
too. Namely, $m_A^2=g^2(|\vev{H^c}|^2+(4/3)
|\vev{\phi}|^2)$ and $m_{W_{\rm R}}^2=g^2
(|\vev{H^c}|^2+2|\vev{\phi}|^2)$. Calculating the
full mass spectrum of the model in the
appropriate SUSY vacuum, one finds that there are
fields acquiring mass of order $m$ and others
that acquire mass of order $g|\vev{H^c}|$. The
presence of cosmic strings
has forced the magnitude of the VEV of the fields
$H^c$, $\bar{H}^c$ in the SUSY vacuum to be in
the range $(1.85-3.21)\times 10^{15}\units{GeV}$
(for $f_{10}=0.02-0.18$), which is about an order
of magnitude below the SUSY GUT scale.
Furthermore, for all the values of the model
parameters encountered here, the highest mass
scale of the model in the SUSY vacuum is
$m_{A^\perp}=\sqrt{5/2}\,g|\vev{H^c}|$. So, we set
this scale equal to the unification scale $M_x$.
From all the above, it is evident that the great
desert hypothesis is not satisfied in this model
and the simple SUSY unification of the gauge
coupling constants is spoiled.

\begin{figure}[tp]
\centering
\includegraphics[width=\linewidth]{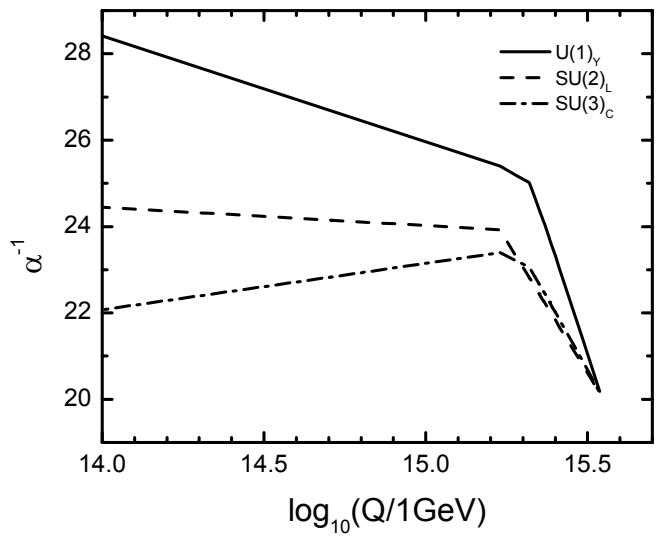}
\caption{Gauge coupling constant unification in
our model for semi-shifted hybrid inflation in
the case of minimal SUGRA for
$v=-\gamma m/4\kappa\lambda$,
$\gamma/2\lambda=1$, and $f_{10}=0.10$. The
parameter $\alpha$ represents the three running
SM fine structure constants as indicated and $Q$
is the running mass scale.}
\label{fig:gauge}
\end{figure}

\par
One can easily see that, although there exist
many fields with ${\rm SU}(3)_c$ and
${\rm U(1)_Y}$ quantum numbers which can acquire
heavy masses below the unification scale and,
thus, affect the running of the corresponding
gauge coupling constants, the only heavy fields
with $\rm{SU(2)_L}$ quantum numbers are $h'$ and
$\bar{h}'$ belonging to the (15,2,2)
representation (see Ref.~\cite{quasi}). However,
these fields affect equally the running of the
${\rm U(1)_Y}$ gauge coupling constant and,
consequently, cannot help us much in achieving
gauge unification. We, therefore, assume that
their masses are close to $M_x$ so that they do
not contribute to the renormalization group
running. As a consequence of these facts, the
$\rm{SU(2)_L}$ gauge coupling constant fails to
unify with the other gauge coupling constants.
One is, thus, forced to consider the inclusion of
some extra fields. There is a good choice using a
single extra field, namely a superfield $f$
belonging to the (15,3,1) representation. This
field affects mainly the running of the
$\rm{SU(2)_L}$ gauge coupling constant. If we
require that $f$ has charge $1/2$ under the
global ${\rm U(1)}$ R symmetry, then the only
renormalizable superpotential term in which this
field is allowed to participate is a mass term of
the form $\frac{1}{2}m_{f}f^2$. One can then tune
the new mass parameter $m_{f}$, along with the
mass $m$, so as to achieve unification of the
gauge coupling constants. In contrast to
Ref.~\cite{stsm}, we will not include here the
superpotential term $\phi^2\pb$ allowed
\cite{quasi} by the symmetries of the model
since, as it turns out, it is not so useful in
the present case. We will assume that the
corresponding coupling constant is negligible.

\par
\begin{figure}[tp]
\centering
\includegraphics[width=\linewidth]{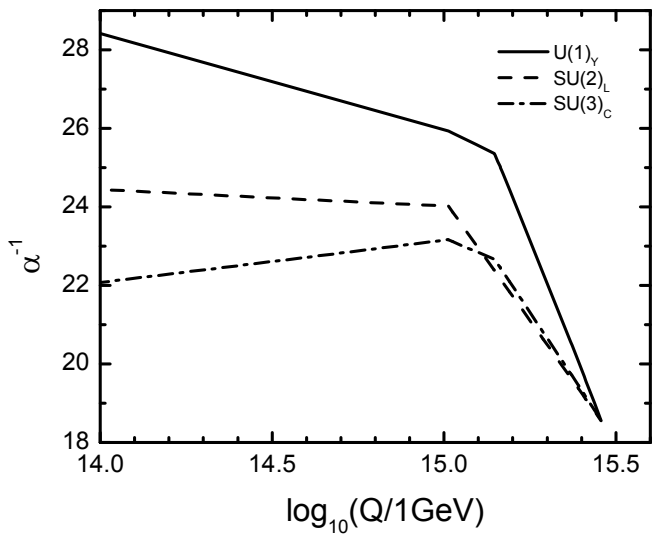}
\caption{Gauge coupling constant unification in
our model for semi-shifted hybrid inflation in
the case of minimal SUGRA for
$v=-\gamma m/4\kappa\lambda$,
$\gamma/2\lambda=1$, and $n_{\rm s}=1$. Same
notation as in Fig.~\ref{fig:gauge}.}
\label{fig:HZgauge}
\end{figure}

We have implemented a code that is built on top
of the {\tt SOFTSUSY} code of
Ref.~\cite{allanach} and performs the running of
the gauge coupling constants at two loops. We
have incorporated six mass thresholds
below the unification scale $M_x$, namely $m_f$,
$m$, $[m^2+(4/3)\la^2|\vev{H^c}|^2]^{1/2}$,
$[m^2+2\la^2|\vev{H^c}|^2]^{1/2}$,
$g[|\vev{H^c}|^2+(4/3)|\vev{\phi}|^2]^{1/2}$,
and $g[|\vev{H^c}|^2+2|\vev{\phi}|^2]^{1/2}$.
In Fig.~\ref{fig:gauge}, we present the
unification of the SM gauge coupling constants in
the $f_{10}=0.10$ case. We deduce that gauge
unification is achieved for
$m_f\simeq 1.69\times 10^{15}\units{GeV}$ and
$m\simeq 2.085\times 10^{15}\units{GeV}$ with
the values of the other parameters of the model
being $n_{\rm s}\simeq 1.0254$,
$M\simeq 2.53\times 10^{15}\units{GeV}$,
$\gamma\simeq 0.515$, $\la\simeq 0.2575$,
$\ka\simeq 0.713$,
$\si_Q\simeq 2.5\times 10^{17}\units{GeV}$,
$\si_f\simeq 4.5\times 10^{16}\units{GeV}$,
$N_Q\simeq 54.2$,
$dn_{\rm s}/d\ln k\simeq -0.8\times 10^{-6}$,
$r\simeq 1.5\ten{-5}$, and the ratio
$\si_f/\si_c\simeq 6.5$. The GUT gauge coupling
constant turns out to be $g\simeq 0.789$ and the
unification scale
$M_x\simeq 3.45\times 10^{15}\units{GeV}$.
In the HZ case (i.e. for
$n_{\rm s}=1$), gauge unification is achieved
for $m_f\simeq 1.025\times 10^{15}\units{GeV}$
and $m\simeq 1.40\times 10^{15}\units{GeV}$ (see
Fig.~\ref{fig:HZgauge}), which corresponds to
$f_{10}\simeq 0.039$,
$M\simeq 1.68\times 10^{15}\units{GeV}$,
$\gamma\simeq 0.367$, $\la\simeq 0.1835$,
$\ka\simeq 0.721$,
$\si_Q\simeq 1.5\times 10^{17}\units{GeV}$,
$\si_f\simeq 3.4\times 10^{16}\units{GeV}$,
$N_Q\simeq 53.9$,
$dn_{\rm s}/d\ln k\simeq -0.2\times 10^{-6}$,
$r\simeq 0.3\times 10^{-5}$,
$\si_f/\si_c\simeq 6.3$, $g\simeq 0.823$, and
$M_x\simeq 2.865\times 10^{15}\units{GeV}$.
Note that the unification scale $M_x$ turns out
to be somewhat small. This fact,
however, does not lead to unacceptably fast
proton decay since the relevant diagrams are
suppressed by large factors (for details, see
Ref.~\cite{shift}).

\section{Conclusions}
\label{sec:conclusions}

\par
It has been shown that the extension of the SUSY
PS model which has been introduced in
Ref.~\cite{quasi} in order to solve the $b$-quark
mass problem in SUSY GUTs with exact asymptotic
Yukawa unification, such as the simplest SUSY PS
model, universal boundary conditions and $\mu>0$
is a very fruitful framework for constructing
hybrid inflationary models. Indeed, it has been
demonstrated that this model automatically and
naturally leads to the so-called new shifted and
new smooth hybrid inflationary scenarios,
which are based only on renormalizable
superpotential terms and avoid the cosmological
disaster from a possible overproduction of PS
magnetic monopoles at the termination of
inflation. These variants of SUSY hybrid
inflation, however, yield, in the context of
minimal SUGRA, values of the spectral index which
lie above the range allowed by the recent CMB
data. It is quite remarkable that this problem
can also be resolved within the same extended
SUSY PS model by considering a two-stage
inflationary scenario without the need of
non-minimal terms in the K\"{a}hler potential.

\par
Here, we have presented an alternative
inflationary scenario which can also be naturally
realized within the same PS model using the same
renormalizable superpotential terms as in the
previous inflationary scenarios and is
compatible with the recent data within minimal
SUGRA. This scenario incorporates cosmic strings
produced at the end of inflation. Our PS model,
in a certain range of parameters, possesses a
semi-shifted classically flat direction on which
the PS gauge group is spontaneously broken only
to $G_{\rm SM}\times{\rm U(1)_{B-L}}$. This
direction acquires a slope from one-loop
radiative corrections originating from the SUSY
breaking caused by the non-zero potential energy
density on this trajectory. Therefore, it can be
used as an inflationary path. We coined
the name semi-shifted hybrid inflation for the
resulting inflationary scenario. As it turns out,
inflation terminates by violating the slow-roll
conditions well before the system reaches the
critical point of the semi-shifted path.
Subsequently, the system crosses the critical
point where the semi-shifted trajectory is
destabilized and the spontaneous breaking of
the ${\rm U(1)_{B-L}}$ gauge symmetry takes
place. As a result of this breaking, a network of
local cosmic strings, which can contribute to the
primordial curvature perturbation, is produced.

\par
It is known that, in the presence of a network of
cosmic strings, the present CMB data can easily
become compatible with
values of the spectral index which are close to
unity or even exceed it. We use a recent fit
\cite{bevis1} to CMB and SDSS data which is based
on field-theory simulations of a dynamical
network of local cosmic strings. For the
power-law $\Lambda$CDM cosmological model, this
fit implies that, at $95\%$ c.l., the spectral
index is $n_{\rm s}=0.94-1.06$ and the fractional
contribution of cosmic strings to the temperature
power spectrum at $\ell=10$ is
$f_{10}=0.02-0.18$. Our
numerical results show that the semi-shifted
hybrid inflationary model can easily become
compatible with this fit without the need of
non-minimal terms in the K\"{a}hler potential or
a subsequent second stage of inflation. Taking
into account the constraints from the unification
of the gauge coupling
constants, we have found that, for a
certain choice of parameters, the model yields
$f_{10}\simeq 0.039$ in the HZ case
(i.e. for $n_{\rm s}=1$) and
$n_{\rm s}\simeq 1.0254$ for the best-fit value
of $f_{10}$ ($=0.10$). Spectral indices
which are lower than about $0.98$ cannot be
obtained. So, the model shows a slight
preference to blue spectra. The cosmological
disaster from the possible overproduction of PS
magnetic monopoles is avoided since there is no
production of such monopoles at the end of the
semi-shifted hybrid inflation.

\section*{ACKNOWLEDGEMENTS}

\par
We thank the authors of Ref.~\cite{bevis1} for
sharing with us some details of their analysis.
This work was supported in part by the European
Commission under the Research and Training
Network contracts MRTN-CT-2004-503369 and
HPRN-CT-2006-035863. It was also supported in
part by the Greek Ministry of Education and
Religion and the Operational Program for
Education and Initial Vocational Training
(EPEAEK) ``Pythagoras''.

\def\ijmp#1#2#3{{Int. Jour. Mod. Phys.}
{\bf #1},~#3~(#2)}
\def\plb#1#2#3{{Phys. Lett. B }{\bf #1},~#3~(#2)}
\def\zpc#1#2#3{{Z. Phys. C }{\bf #1},~#3~(#2)}
\def\prl#1#2#3{{Phys. Rev. Lett.}
{\bf #1},~#3~(#2)}
\def\rmp#1#2#3{{Rev. Mod. Phys.}
{\bf #1},~#3~(#2)}
\def\prep#1#2#3{{Phys. Rep. }{\bf #1},~#3~(#2)}
\def\prd#1#2#3{{Phys. Rev. D }{\bf #1},~#3~(#2)}
\def\npb#1#2#3{{Nucl. Phys. }{\bf B#1},~#3~(#2)}
\def\npps#1#2#3{{Nucl. Phys. B (Proc. Sup.)}
{\bf #1},~#3~(#2)}
\def\mpl#1#2#3{{Mod. Phys. Lett.}
{\bf #1},~#3~(#2)}
\def\arnps#1#2#3{{Annu. Rev. Nucl. Part. Sci.}
{\bf #1},~#3~(#2)}
\def\sjnp#1#2#3{{Sov. J. Nucl. Phys.}
{\bf #1},~#3~(#2)}
\def\jetp#1#2#3{{JETP Lett. }{\bf #1},~#3~(#2)}
\def\app#1#2#3{{Acta Phys. Polon.}
{\bf #1},~#3~(#2)}
\def\rnc#1#2#3{{Riv. Nuovo Cim.}
{\bf #1},~#3~(#2)}
\def\ap#1#2#3{{Ann. Phys. }{\bf #1},~#3~(#2)}
\def\ptp#1#2#3{{Prog. Theor. Phys.}
{\bf #1},~#3~(#2)}
\def\apjl#1#2#3{{Astrophys. J. Lett.}
{\bf #1},~#3~(#2)}
\def\n#1#2#3{{Nature }{\bf #1},~#3~(#2)}
\def\apj#1#2#3{{Astrophys. J.}
{\bf #1},~#3~(#2)}
\def\anj#1#2#3{{Astron. J. }{\bf #1},~#3~(#2)}
\def\apjs#1#2#3{{Astrophys. J. Suppl.}
{\bf #1},~#3~(#2)}
\def\mnras#1#2#3{{MNRAS }{\bf #1},~#3~(#2)}
\def\grg#1#2#3{{Gen. Rel. Grav.}
{\bf #1},~#3~(#2)}
\def\s#1#2#3{{Science }{\bf #1},~#3~(#2)}
\def\baas#1#2#3{{Bull. Am. Astron. Soc.}
{\bf #1},~#3~(#2)}
\def\ibid#1#2#3{{\it ibid. }{\bf #1},~#3~(#2)}
\def\cpc#1#2#3{{Comput. Phys. Commun.}
{\bf #1},~#3~(#2)}
\def\astp#1#2#3{{Astropart. Phys.}
{\bf #1},~#3~(#2)}
\def\epjc#1#2#3{{Eur. Phys. J. C}
{\bf #1},~#3~(#2)}
\def\nima#1#2#3{{Nucl. Instrum. Meth. A}
{\bf #1},~#3~(#2)}
\def\jhep#1#2#3{{J. High Energy Phys.}
#1~(#2)~#3}
\def\lnp#1#2#3{{Lect. Notes Phys.}
{\bf #1},~#3~(#2)}
\def\appb#1#2#3{{Acta Phys. Polon. B}
{\bf #1},~#3~(#2)}
\def\njp#1#2#3{{New J. Phys.}
{\bf #1},~#3~(#2)}
\def\pl#1#2#3{{Phys. Lett. }{\bf #1B},~#3~(#2)}
\def\jcap#1#2#3{{J. Cosmol. Astropart. Phys.}
#1~(#2)~#3}
\def\mpla#1#2#3{{Mod. Phys. Lett. A}
{\bf #1},~#3~(#2)}
\def\jpcs#1#2#3{{J. Phys. Conf. Ser.}
{\bf #1},~#3~(#2)}
\def\jetpl#1#2#3{{JETP Lett.}
{\bf #1},~#3~(#2)}
\def\rpp#1#2#3{{Rep. Prog. Phys.}
{\bf #1},~#3~(#2)}
\def\pl#1#2#3{{Phys. Lett.}
{\bf #1},~#3~(#2)}
\def\cpc#1#2#3{{Comput. Phys. Commun.}
{\bf #1},~#3~(#2)}


\begin{thebibliography}{99}

\bibitem{inflation}
A.H. Guth, \prd{23}{1981}{347}.

\bibitem{lectures}
G. Lazarides, \lnp{592}{2002}{351},
arXiv: hep-ph/0111328;
\jpcs{53}{2006}{528}, arXiv:hep-ph/0607032.

\bibitem{linde}
A.D. Linde, \prd{49}{1994}{748}.

\bibitem{cop}
E.J. Copeland, A.R. Liddle, D.H. Lyth,
E.D. Stewart, and D. Wands,
\prd{49}{1994}{6410}.

\bibitem{dss}
G.R. Dvali, Q. Shafi, and R.K. Schaefer,
\prl{73}{1994}{1886};
G. Lazarides, R.K. Schaefer, and Q. Shafi,
\prd{56}{1997}{1324}.

\bibitem{monopole}
G.'t Hooft, \npb{79}{1974}{276};
A.M. Polyakov, \jetpl{20}{1974}{194};
J.P. Preskill, \prl{43}{1979}{1365};
G. Lazarides, Q. Shafi, and W.P. Trower,
\ibid{49}{1982}{1756}.

\bibitem{smooth1}
G. Lazarides and C. Panagiotakopoulos,
\prd{52}{1995}{R559}.

\bibitem{smooth2}
G. Lazarides, C. Panagiotakopoulos, and
N.D. Vlachos, \prd{54}{1996}{1369};
R. Jeannerot, S. Khalil, and G. Lazarides,
\plb{506}{2001}{344}.

\bibitem{shift}
R. Jeannerot, S. Khalil, G. Lazarides, and
Q. Shafi, \jhep{10}{2000}{012}.

\bibitem{talks}
G. Lazarides, arXiv:hep-ph/0011130;
R. Jeannerot, S. Khalil, and G. Lazarides,
arXiv:hep-ph/0106035.

\bibitem{nsmooth}
G. Lazarides and A. Vamvasakis,
Phys.\ Rev.\ D {\bf 76}, 083507 (2007).

\bibitem{nshift}
R. Jeannerot, S. Khalil, and G. Lazarides,
\jhep{07}{2002}{069}.

\bibitem{pati}
J.C. Pati and A. Salam, \prd{10}{1974}{275}.

\bibitem{magg}
G. Lazarides, M. Magg, and Q. Shafi,
\pl{97B}{1980}{87}.

\bibitem{quasi}
M.E. Gomez, G. Lazarides, and C. Pallis,
\npb{638}{2002}{165}.

\bibitem{quasitalks}
G. Lazarides and C. Pallis, arXiv:hep-ph/0404266;
arXiv: hep-ph/0406081.

\bibitem{als}
B. Ananthanarayan, G. Lazarides, and
Q. Shafi, \prd{44}{1991}{1613};
\plb{300}{1993}{245}.

\bibitem{hw}
G. Lazarides and C. Panagiotakopoulos,
\plb{337}{1994}{90};
S. Khalil, G. Lazarides, and C. Pallis,
\ibid{508}{2001}{327}.

\bibitem{hall}
R. Hempfling, \prd{49}{1994}{6168};
L.J. Hall, R. Rattazzi, and U. Sarid,
\ibid{50}{1994}{7048}.

\bibitem{wetterich}
G. Lazarides, Q. Shafi, and C. Wetterich,
\npb{181}{1981}{287};
G. Lazarides and Q. Shafi,
\ibid{B350}{1991}{179}.

\bibitem{wmap} D.N. Spergel {\it et al.},
\apjs{170}{2007}{377};
E. Komatsu {\it et al.}, arXiv:0803.0547.

\bibitem{newns}
K.M. Huffenberger, H.K. Eriksen, F.K. Hansen,
A.J. Banday, and K.M. Gorski, arXiv:0710.1873;
C.L. Rei-chardt {\it et al.}, arXiv:0801.1491.

\bibitem{senoguz}
V.N. \c{S}eno\u{g}uz and Q. Shafi,
\plb{567}{2003}{79}.

\bibitem{lofti}
L. Boubekeur and D. Lyth, \jcap{07}{2005}{010}.

\bibitem{king}
M. Bastero-Gil, S.F. King, and Q. Shafi,
\plb{651}{2007}{345}.

\bibitem{rehman}
M. ur Rehman, V.N. \c{S}eno\u{g}uz, and Q. Shafi,
\prd{75}{2007}{043522}.

\bibitem{mhin}
G. Lazarides and C. Pallis, \plb{651}{2007}{216};
G. Lazarides, arXiv:0706.1436.

\bibitem{stsm}
G. Lazarides and A. Vamvasakis, Phys. Rev. D
{\bf 76}, 123514 (2007).

\bibitem{string}
H.B. Nielsen and P. Olesen,
\npb{61}{1973}{45};
G. Lazarides, Q. Shafi, and T.F. Walsh,
{\it ibid.} {\bf B195}, 157 (1982);
T.W.B. Kibble, G. Lazarides, and Q. Shafi,
\pl{113B}{1982}{237}.

\bibitem{vilenkin}
A. Vilenkin and E.P.S. Shellard, {\it Cosmic
Strings and Other Topological Defects} (Cambridge
University Press, Cambridge, U.K., 1994);
M.B. Hindmarsh and T.W.B. Kibble,
\rpp{58}{1995}{477}.

\bibitem{rc}
G.R. Dvali, Q. Shafi, and R.K. Schaefer,
\prl{73}{1994}{1886};
G. Lazarides, R.K. Schaefer, and Q. Shafi,
\prd{56}{1997}{1324}.

\bibitem{battye}
R.A. Battye, B. Garbrecht, and A. Moss,
\jcap{09}{2006}{007}.

\bibitem{sdss}
M. Tegmark {\it et al.}, \prd{74}{2006}{123507}.

\bibitem{bevis1}
N. Bevis, M. Hindmarsh, M. Kunz, and
J. Urrestilla, \prl{100}{2008}{021301}.

\bibitem{bevis2}
N. Bevis, M. Hindmarsh, M. Kunz, and
J. Urrestilla, \prd{75}{2007}{065015};
\ibid{76}{2007}{043005}.

\bibitem{laz1}
G. Lazarides, C. Panagiotakopoulos, and Q. Shafi,
\prl{56}{1986}{432}.

\bibitem{laz2}
N. Ganoulis, G. Lazarides, and Q. Shafi,
\npb{323}{1989}{374}.

\bibitem{sstad}
V.N. \c{S}eno\u{g}uz and Q. Shafi,
Phys. Rev. D {\bf 71}, 043514 (2005);
arXiv:hep-ph/0512170.

\bibitem{ColemanWeinberg}
S.R. Coleman and E. Weinberg, Phys. Rev. D
{\bf 7}, 1888 (1973).

\bibitem{review}
B.A. Bassett, S. Tsujikawa, and D. Wands,
\rmp{78}{2006}{537}.

\bibitem{gravitino}
M.Yu. Khlopov and A.D. Linde,
Phys. Lett. {\bf 138B}, 265 (1984);
J. Ellis, J.E. Kim, and D. Nanopoulos,
\ibid{145B}{1984}{181};
J.R. Ellis, D.V. Nanopoulos, and S. Sarkar,
\npb{259}{1985}{175}.

\bibitem{private}
N. Bevis, M. Hindmarsh, M. Kunz, and
J. Urrestilla, private communication.

\bibitem{superconduct}
N. Ganoulis and G. Lazarides,
Phys. Rev. D {\bf 38}, 547 (1988);
\npb{316}{1989}{443}.

\bibitem{allanach}
B.C. Allanach,
Comput. Phys. Commun. {\bf 143}, 305 (20 02).

\end{thebibliography}
\end{document}